\documentclass[5p,onecolumn]{elsarticle}

\usepackage{hyperref}
\usepackage[utf8]{inputenc}
\usepackage[T1]{fontenc}
\usepackage{lmodern}
\usepackage{microtype}
\usepackage{csquotes}
\usepackage[english]{babel}
\usepackage{amsmath,amssymb,amsfonts}
\usepackage{graphicx}
\usepackage{adjustbox}
\usepackage{textcomp}
\usepackage{xcolor}
\usepackage{xspace}
\usepackage{tcolorbox}
\usepackage{csvsimple}
\usepackage{booktabs}
\usepackage{arydshln}
\usepackage{siunitx}
\usepackage{balance}
\usepackage{url}
\usepackage{tikz}
\usepackage{fontawesome}
\usepackage{todonotes}
\usepackage{braket}
\usepackage{cleveref}
\usepackage{listings}
\usepackage{adjustbox}
\usepackage{color, colortbl}

\journal{Journal of Systems and Software}

\newcommand{\ie}{i.e.,\xspace}
\newcommand{\eg}{e.g.,\xspace}

\newcommand{\etal}{\emph{et al.}\xspace}

\usepackage{enumitem}

\definecolor{gray50}{gray}{.5}
\definecolor{gray40}{gray}{.6}
\definecolor{gray30}{gray}{.7}
\definecolor{gray20}{gray}{.8}
\definecolor{gray10}{gray}{.9}
\definecolor{gray05}{gray}{.95}

\newlength\Linewidth
\def\findlength{\setlength\Linewidth\linewidth
	\addtolength\Linewidth{-4\fboxrule}
	\addtolength\Linewidth{-3\fboxsep}
}

\newenvironment{rqbox}{\par\begingroup
	\setlength{\fboxsep}{5pt}\findlength
	\setbox0=\vbox\bgroup\noindent
	\hsize=0.95\linewidth
	\begin{minipage}{0.95\linewidth}\normalsize}
	{\end{minipage}\egroup
	\textcolor{gray20}{\fboxsep1.5pt\fbox
		{\fboxsep5pt\colorbox{gray05}{\normalcolor\box0}}}
	\endgroup\par\noindent
	\normalcolor\ignorespacesafterend}

\definecolor{pblue}{rgb}{0.13,0.13,1}
\definecolor{pgreen}{rgb}{0,0.5,0}
\definecolor{pred}{rgb}{0.9,0,0}
\definecolor{pgrey}{rgb}{0.46,0.45,0.48}
\definecolor{codebackground}{rgb}{0.95, 0.95, 0.92}
\definecolor{arsenic}{rgb}{0.23, 0.27, 0.29}

\newtcolorbox{summarybox}[1][RQ]{
	colback=gray!5,
	colframe=arsenic,
	boxrule=0.4mm,
	left=1.5mm,
	right=1.5mm,
	top=1.5mm,
	bottom=1.5mm,
	fonttitle=\bfseries,
	title=Main findings for #1
}

\newcommand{\examplebox}[1]{ %
	\vspace{5pt} %
	\noindent\fcolorbox{arsenic}{gray10}{%
		\parbox{0.97\linewidth}{% 
			\textbf{Discussion - Example.} #1
		}%
	}%
	\vspace{5pt} %
}%

\definecolor{aliceblue}{rgb}{0.94, 0.97, 1.0}

\RequirePackage{fontawesome}
\newcommand{\keyfindingsone}[1]{ %
	\vspace{5pt} %
	\noindent\fcolorbox{arsenic}{aliceblue}{%
		\parbox{0.97\linewidth}{% 
			\textbf{\faKey\ Take Away Message.} #1 %
		}%
	}%
	\vspace{5pt} %
}%

\newcommand{\qiskit}{\textsc{Qiskit}\xspace}
\newcommand{\cirq}{\textsc{Cirq}\xspace}
\newcommand{\qsharp}{\textsc{Q$\sharp$}\xspace}
\newcommand{\totalrepos}{731\xspace}

\definecolor{codegreen}{rgb}{0,0.6,0}
\definecolor{codegray}{rgb}{0.5,0.5,0.5}
\definecolor{codepurple}{rgb}{0.58,0,0.82}
\definecolor{backcolour}{rgb}{0.95,0.95,0.92}
\definecolor{Gray}{gray}{0.93}

\lstdefinestyle{mystyle}{
	backgroundcolor=\color{backcolour},   
	commentstyle=\color{codegreen},
	keywordstyle=\color{magenta},
	numberstyle=\tiny\color{codegray},
	stringstyle=\color{codepurple},
	basicstyle=\ttfamily\footnotesize,
	breakatwhitespace=false,         
	breaklines=true,                 
	captionpos=b,                    
	keepspaces=true,                 
	numbers=left,                    
	numbersep=5pt,                  
	showspaces=false,                
	showstringspaces=false,
	showtabs=false,                  
	tabsize=2
}

\newcounter{Finding}
\stepcounter{Finding}

\newcommand{\revised}[1]{\textcolor{black}{#1}}
\newcommand{\revtwo}[1]{\textcolor{black}{#1}}

\begin{document}

\begin{frontmatter}

\title{Software Engineering for Quantum Programming: How Far Are We?}

%% Group authors per affiliation:
\author[mymainaddress]{Manuel De Stefano}
\cortext[mycorrespondingauthor]{Corresponding author}
\ead{madestefano@unisa.it}

\author[secondaddress]{Fabiano Pecorelli}
\ead{fabiano.pecoreli@tuni.fi}

\author[mymainaddress]{Dario Di Nucci}
\ead{ddinucci@unisa.it}

\author[mymainaddress]{Fabio Palomba}
\ead{fpalomba@unisa.it}

\author[mymainaddress]{Andrea De Lucia}
\ead{adelucia@unisa.it}

\address[mymainaddress]{SeSa Lab - University of Salerno, Italy}
\address[secondaddress]{Tampere University, Finland}

\begin{abstract}
	% !TeX spellcheck = en_US
Quantum computing is no longer only a scientific interest but is rapidly becoming an industrially available technology that can potentially overcome the limits of classical computation. 
Over the last years, all major companies have provided frameworks and programming languages that allow developers to create their quantum applications.
This shift has led to the definition of a new discipline called \emph{quantum software engineering}, which is demanded to define novel methods for engineering large-scale quantum applications. 
While the research community is successfully embracing this call, we notice a lack of systematic investigations into the state of the practice of quantum programming. 
Understanding the challenges that quantum developers face is vital to precisely define the aims of quantum software engineering.
\revised{Hence, in this paper, we first mine all the \textsc{GitHub} repositories that make use of the most used quantum programming frameworks currently on the market and then conduct coding analysis sessions to produce a taxonomy of the purposes for which quantum technologies are used}.
In the second place, we conduct a survey study that involves the contributors of the considered repositories, which aims to elicit the developers' opinions on the current adoption and challenges of quantum programming. 
\revised{On the one hand, the results highlight that the current adoption of quantum programming is still limited. On the other hand, there are many challenges that the software engineering community should carefully consider: these do not strictly pertain to technical concerns but also socio-technical matters}.
\end{abstract}


\end{frontmatter}

% !TeX spellcheck = en_US
\section{Introduction}
\label{sec:intro}

The dream has come true~\cite{knight2018serious}: several physicians and computer scientists agree that the quantum technology is right around the corner~\cite{knight2018serious,hoare2005grand} and that the 21st century will be recalled as the \emph{``quantum era''}~\cite{piattini2021toward}. Specific mechanic principles such as \emph{superposition}, \ie quantum objects may assume different states at the same time, and \emph{entanglement}, \ie quantum objects may be deeply connected without any direct physical interaction, promise to revolutionize program computation compared to classical computers~\cite{mueck2017quantum}. Quantum computers could eventually lead to resolving NP-complete problems \cite{aaronson2005guest,ohya2008new}---often referred to as \emph{quantum supremacy}~\cite{arute2019quantum}, namely the point in time when a programmable quantum device would be able to solve problems that no classical computer can solve in any feasible amount of time. 

For this reason, all major software companies, like \textsc{IBM} and \textsc{Google}, are currently investing hundreds of millions of dollars every year to produce novel hardware and software technologies that can support the execution of quantum programs.\footnote{Boston Consulting Group report: \url{shorturl.at/mINWY}.} For instance, \textsc{IBM Quantum}\footnote{\textsc{IBM Quantum}: \url{https://www.ibm.com/quantum-computing/}.} has developed its programming framework, which allows developers to design, implement, and execute quantum applications on cloud-based quantum computers. Companies and researchers have also been developing several quantum programming languages~\cite{omer2003qcl,qsharp,altenkirch2005qml} and development toolkits~\cite{aleksandrowicz2019qiskit,broughton2020tensorflow,steiger2018projectq} that provide developers with off-the-shelf instruments and APIs to create quantum programs. 

While there have been already several promising applications of quantum programming to the resolution of various problems in the fields of machine learning~\cite{biamonte2017quantum}, optimization~\cite{guerreschi2017practical}, cryptography~\cite{mailloux2016post}, and chemistry~\cite{reiher2017elucidating}, the development of large-scale quantum software seems to be still far from being a reality. In this respect, researchers such as Piattini \etal~\cite{piattini2021toward,piattini2020talavera,piattini2020quantum}, Moguel \etal~\cite{moguel2020roadmap}, and Zhao~\cite{zhao2020quantum} advocated the need for a brand new scientific discipline able to rework and extend the classical software engineering into the quantum domain. This new field that should enable developers to design and develop quantum programs with the same confidence as classical programs is what we call \emph{quantum software engineering} (QSE)~\cite{zhao2020quantum}. 

In response to the quantum software engineering call, our research community has proposed thematic workshops, like \textsc{Q-Se}\footnote{The \textsc{Q-Se} workshop: \url{https://q-se.github.io/qse2021/}.}  \revised{and \textsc{Q-SET},\footnote{The \textsc{Q-SET} workshop: \url{https://quset.github.io/qset2021/}}} as well as devising novel processes~\cite{barbosa2020software}, modeling techniques~\cite{gemeinhardt2021towards}, and debugging mechanisms~\cite{li2020projection}.
Recognizing the initial effort spent by the research community, we notice a notable lack of empirical investigations to provide a more comprehensive overview of the \emph{current state of the practice} on quantum software engineering. 
Therefore, there is a need to analyze how quantum programming is currently used and the key software engineering challenges developers face when programming quantum programs. An improved understanding of these critical aspects might shed light on the limitations of the state of the art and drive the next research steps that our community should invest in. 

\revised{To the best of our knowledge, El Aoun \etal~\cite{Elaoun2021icsme} have been the first to work along these lines. 
They conducted an empirical study on the questions asked by quantum developers on \textsc{Stack Exchange} forums, as well as the issues reported on \textsc{GitHub}. 
The authors performed qualitative coding analyses and automated topic modeling to uncover the topics in quantum software engineering-related posts and issue reports. 
According to the reported results, knowledge of quantum theory and quantum program comprehension represent key engineering aspects hindering the developer's capabilities to develop quantum programs.}

\revised{In this paper, we aim at complementing and extending the work by El Aoun \etal~\cite{Elaoun2021icsme}. Rather than analyzing the traces left by developers over forums and issues, we approach the understanding of the current state of the practice on quantum programming by proposing a two-step investigation that includes a mining software repository analysis and a survey study. The ultimate goal of our study is to assess how far quantum software engineering is from effectively supporting developers in practice.}

\revised{In particular, we first mine all the \textsc{GitHub} repositories that employ the three most widely used quantum programming frameworks implementing the quantum logic gate model, \ie \qiskit~\cite{aleksandrowicz2019qiskit}, \cirq~\cite{cirq_developers_2021_4750446}, and \qsharp~\cite{microsoft_quantum}. Our choice to focus on the emerging technologies enabling the universal quantum gate model is driven by the fact that this can be applied to a broader range of problems if compared to other models, \eg quantum annealing \cite{finnila1994quantum} or quantum adiabatic \cite{farhi2000quantum}.}
We conduct content analysis sessions~\cite{lidwell2010universal} to elicit a taxonomy of \revised{purposes for which quantum programming is used}. 
Afterwards, we run a survey among the contributors of the mined repositories, asking about their opinions and perspectives on the current adoption of quantum programming and the key challenges they face when engaging with the currently available quantum technologies.  

On the one hand, our repository analysis highlights that quantum programming is currently mainly used for didactic purposes or for curiosity to experiment with quantum technologies. On the other hand, the survey study identifies five significant challenges related to the comprehension of quantum programs, the hardness of setting up hardware and software infrastructures, the implementation and code quality issues, the difficulty of building a quantum developer's community, and, perhaps more importantly, the lack of realism of the current quantum applications. 

\revised{Based on these observations, we conclude that the road ahead for quantum software engineering is still long and concerns technical and socio-technical matters.} 

\smallskip
\noindent \textbf{Structure of the paper.} \revised{\Cref{sec:background} overviews the basic concepts behind quantum computing and programming, while \Cref{sec:related} summarizes the existing literature. \Cref{sec:design} elaborates on the research questions, the context of the study, and the methodology of our empirical study. In Sections \ref{sec:rq1} and \ref{sec:taxonomy} we analyze the results addressing the two research questions of the study. \Cref{sec:implications} discusses the main findings and the implications that they have for researchers, practitioners, and tool vendors. The threats to the validity of the study are reported and discussed in \Cref{sec:ttv}. Finally, \Cref{sec:conclusion} concludes the paper and outlines our future research agenda.}

% !TeX spellcheck = en_US
\section{\revised{Background}} \label{sec:background}

\revised{In the following, we describe the basics for understanding the universal gate model of quantum computing \cite{barenco1995elementary, feynman2017quantum}.}
%concepts of quantum computing and programming. %Afterwards, we report on the related literature, highlighting the main limitations that our work aims at addressing.
%The background refers to the universal gate model of quantum computing \cite{barenco1995elementary, feynman2017quantum}.}
%Quantum Annealing \cite{finnila1994quantum} and Quantum Adiabatic \cite{farhi2000quantum} Models and their implementation are out of the scope of this paper.}

\subsection{Quantum Bits}
In classical programming, the base unit of information is the bit, which can assume only binary values.
In quantum computing, the base unit of information is the \emph{quantum} bit (\ie \emph{qubit} or \emph{qbit}).
A qubit differs from a classical bit since its state is a linear combination of two bases in the quantum space, represented by a bidimensional vector \cite{kaye2007introduction}.
Thus, the computational basis of the qubit is defined as:
\begin{equation}
        \ket{0} = \begin{bmatrix}
        1 \\
        0
    \end{bmatrix} \;
    \ket{1} = \begin{bmatrix}
        0 \\
        1
    \end{bmatrix}
\end{equation}
Then, a generic qubit $\ket{q}$ can be represented as:
\begin{equation}
    \ket{q} = \alpha\ket{0} + \beta\ket{1}
\end{equation}
where $\alpha$ and $\beta$ are complex numbers subject to the \emph{normalization condition}, \ie $|\alpha|^2 + |\beta|^2 = 1 $.
This is necessary since $|\alpha|^2$ and $|\beta|^2$ indicates the probability of the qubit to be either in state 0 or 1.
The fact that the qubit can be found in one of these states with a certain probability is called \emph{superposition} \cite{kaye2007introduction}.
In other words, a quantum computer consisting of Qubits is in many different states at the same time.

\subsection{Quantum Gates}
In classical logical circuits, the state of a bit can be changed by applying a gate. The same concept applies to quantum circuits. 
In this case, such gates can be applied on a single (\ie for the transition of a single quantum state) or multiple qubits (\ie for the transition of multiple quantum states). 
The number of inputs and outputs \revised{of a gate} should be equal to make the operation reversible.
A \emph{single qubit gate} is represented by a squared bidimensional matrix. 
The resulting quantum state is determined by multiplying the quantum state vector with the matrix.
An example of a single qubit gate is the $NOT$ gate, denoted in the following.

\begin{equation}
    NOT = \begin{bmatrix}
        0 & 1\\
        1 & 0
    \end{bmatrix}
\end{equation}
Applying a $NOT$ gate to a generic qubit $[\alpha,\beta]$ results in a new state vector whose components are inverted, as shown in the following:
\begin{equation}    
        NOT\begin{bmatrix}
            \alpha\\
            \beta
        \end{bmatrix} = 
        \begin{bmatrix}
            0 & 1\\
            1 & 0
        \end{bmatrix}
        \begin{bmatrix}
            \alpha\\
            \beta
        \end{bmatrix} =
        \begin{bmatrix}
            \beta\\
            \alpha
        \end{bmatrix}
\end{equation}
Single qubit gates are fundamental to achieve the \emph{superposition} condition described above.

\revised{Conditional logic operations are needed to change the state of a qubit given the state of another qubit. They require quantum gates with multiple inputs and outputs, namely \emph{multiple qubits gates}, which are represented as square matrices having a higher dimension.}
An example is the \emph{controlled not} ($CNOT$):
\begin{equation}
    CNOT = \begin{bmatrix}
        1 & 0 & 0 & 0\\
        0 & 1 & 0 & 0\\
        0 & 0 & 0 & 1\\
        0 & 0 & 1 & 0\\
    \end{bmatrix}
\end{equation} 
This operation takes as input two qubits: the control qubit and the target qubit.
When the control qubit has state $\ket{0}$, the target qubit remains unchanged, whereas when the control qubit has state $\ket{1}$, a \textsc{NOT} gate is applied to the target qubit.
Please consider that in both cases, the control qubit remains unchanged.
By leveraging multiple qubit gates, \revised{in particular the CNOT and the H gate}, it is possible to achieve the \emph{entanglement} \cite{kaye2007introduction} among qubits, which is a condition in which the state of a set of qubits cannot be described anymore by considering the state of a single qubit, but considering the system as a whole.
\revised{An entangled system is one whose quantum state cannot be factored as a product of the states of its local constituents; in other words, they are not discrete particles but rather an indivisible totality. 
If two constituents are entangled, one discrete particle cannot be properly defined without addressing the other. The state of a composite system may always be expressed as a sum, or superposition, of products of local constituent states; it is entangled if this sum cannot be represented as a single product term \cite{kaye2007introduction}.
In these cases, changing one qubit will affect all other constituents.
This particular situation can be achieved (for instance), if we put a \textsc{CNOT} gate on a two-qubit registry, and the control qubit is in superposition (to be precise the $ \ket{+} = \frac{1}{\sqrt{2}}(\ket{00} + \ket{11})$ state).
Entanglement plays a crucial role for quantum computing since it is necessary for a quantum algorithm to offer an exponential speed-up over classical computations \cite{jozsa2003role}.
Indeed, changing the state of an entangled qubit will immediately change the state of all paired qubits, speeding up the process.}

\subsection{\revised{Quantum Measurement}}
\revised{The measurement, \ie the process to obtain the information of the qubits, is one of the most important operations appliable to qubits.
Measurements are irreversible and permanently force qubits to certain states (i.e., 0 or 1).
To be more precise, the probability of measuring a state  $\ket{\psi}$ in any qubit state $\ket{x}$ is given by the following equation}:

\begin{equation}
    \revised{p(\ket{x}) = |\braket{x|\psi}|^2}
\end{equation} 

\revised{This action has several implications.
Any qubit in the state $\ket{\psi} = \alpha\ket{0} + \beta\ket{1}$ has $ |\alpha|^2$ probability to be found in the state $\ket{0}$, and $ |\beta|^2$ probability to be found in the state $\ket{1}$.
Measurements alter the magnitudes of $\alpha$ and $\beta$. For instance, if the result of the measurement is $\ket{1}$, $\alpha$ is changed to 0, and $\beta$ is changed to the phase factor $ e^{i\phi }$ that is no longer experimentally accessible. 
If a measurement is performed on an entangled qubit, it may collapse the state of the other entangled qubits.
}

\subsection{Quantum Programming}
Combining qubits and gates, we obtain the \emph{quantum (logical) circuit}, one of the most commonly used and general-purpose quantum computing models \cite{kaye2007introduction}.
This allows to leverage on \emph{superposition} and \emph{entanglement} among qubits and achieve a \emph{theoretical} advantage over classical computers in performing large-scale parallel computation \cite{kaye2007introduction}.

To determine the type of transformation the circuit performs, we need to analyze the structure of quantum circuits, the number and the type of gates, and the interconnection scheme.
The result of a quantum circuit can be read out through \emph{quantum measurements}.

%\begin{adjustbox}{width=1\columnwidth}
	\begin{lstlisting}[language=Python,caption=\qiskit code for the bell state circuit., label=lst:bell] 
		from qiskit import QuantumCircuit
		# Create a quantum circuit with two qubits and classical bits:
		qc = QuantumCircuit(2,2)
		# Apply H-gate to the first:
		qc.h(0)
		# Apply a CNOT:
		qc.cx(0,1)	
    # measure the qubits
    qc.measure(0,0)
    qc.measure(1,1)
	\end{lstlisting}
%\end{adjustbox}
	
Quantum programming is the process of designing and building executable code that a quantum computer will execute to obtain a particular result~\cite{miszczak2012high,ying2016foundations}.
A quantum program contains classical code blocks and quantum components.
\revised{As a result, a typical quantum program contains two types of instructions (or statements), namely quantum and classical statements}. 
On the one hand, classical instructions work with the state of classical bits and apply conditional expressions.
On the other hand, quantum instructions operate on the state of qubits and measure qubit values.
These operations, which can be applied to (\emph{single qubit}) and (\emph{multiple qubits}) and can be reversible, are mainly represented as quantum circuits that manipulate qubit register to perform quantum operations.
At the end of the operation, classical bit registers are used to store \emph{quantum measurements}.

\smallskip
Quantum programming, at the state of the practice, heavily leverages libraries and APIs (\eg \qiskit) to define quantum circuits and run them on quantum machines.
Listing~\ref{lst:bell} shows the code of a simple circuit, which puts the 2 qubits in an entangled state called \textit{bell state}, which is a state in which the only two possible measures can be \texttt{00} or \texttt{11}, with a 50\% probability each.
As can be seen, to assemble a quantum circuit, the class \texttt{QuantumCircuit} is imported from the \qiskit library.
Then, the quantum gates are applied on the qubits invoking some API defined on the same \texttt{QuantumCircuit} class.
% !TeX spellcheck = en_US
\section{\revised{Related Work}} \label{sec:related}
\begin{table*}[t]
	\centering
	\caption{Comparison between the state-of-the art work which explicit challenges in quantum software engineering.} \label{tab:related}
	\resizebox{\textwidth}{!}{
		\begin{tabular}{@{}p{0.2\textwidth}p{0.4\textwidth}p{0.4\textwidth}@{}}
			\toprule
			\bfseries Paper Title & \bfseries Methodology & \bfseries Main Results \\
			\midrule
			Toward a Quantum Software Engineering \cite{piattini2021toward} & Analysis of the impact that the rules defined in the Talavera Manifesto \cite{piattini2020talavera} can have in the fields of software engineering. & Proposal of Some \textit{hot topics} to focus on. In particular, some relevant advice is given to practitioners, researchers and universities.\\ \addlinespace[0.5em] 	
			Quantum Software Engineering: Landscapes and Horizons \cite{zhao2020quantum} & Non-Systematic survey of the current literature regarding quantum software engineering & Definition of a set of challenges divided by software engineering areas, \ie quantum requirements analysis, quantum software design, \revtwo{and quantum software testing.} \\ \addlinespace[0.5em]		
			Understanding Quantum Software Engineering Challenges: An Empirical Study on Stack Exchange Forums and GitHub Issues \cite{Elaoun2021icsme} & Qualitative analysis of QSE-related questions raised by developers on Stack Exchange forums. Automated topic modeling to uncover the QSE-related Stack Exchange posts and GitHub issue reports. & Highlighting some of the more difficult aspects of quantum software engineering that are distinct from traditional software engineering (\eg explaining quantum code).\\ \addlinespace[0.5em]
            \rowcolor{Gray}\bfseries Our Work & \bfseries Combination of mining software repositories and a survey with practitioners. & \bfseries Taxonomies of current usages and challenges, perceived by quantum programmers, of quantum programming.\\
			\bottomrule
		\end{tabular}
	}
\end{table*}

Research in software engineering for quantum programming, or quantum software engineering (QSE), is in its infancy.
During the first International Workshop on Quantum Software Engineering, researchers and practitioners proposed a QSE manifesto, known as the ``Talavera Manifesto", which defines the set of fundamental principles of this new discipline~\cite{piattini2020talavera, piattini2020quantum}.
Some of these principles include (i) agnosticism towards specific quantum technologies, (ii) coexistence of classical and quantum programming, (iii) support for developing and maintaining quantum software.
Since then, several studies have been presented discussing challenges and potential direction in QSE research under various perspectives. Details are reported in \Cref{tab:related}.

Piattini \etal \cite{piattini2021toward} defined a set of topics on which researchers should pay attention. 
In particular, after digression on the ``Golden Eras of Software Engineering'', some priority areas for quantum software engineering were explored.
To be more precise, they pointed out 4 areas of quantum software engineering which need high attention from developers: software design of quantum hybrid systems, testing techniques for quantum programming, quantum programs quality, and re-engineering and modernization toward classical-quantum information systems.
Finally, they provide some useful insights for researchers, practitioners and universities, described in the following.
According to them, researchers should consider that many quantum computer scientist do not have a sufficient knowledge of software engineering techniques, so many errors might be done again, and some expensive ``rediscoveries'' could happen.
Moreover, they think that researchers should not wait until quantum programming languages would be ``stable'' or ``refined'' in order to propose, adapt or develop software engineering quantum techniques, but develop them in parallel with the evolution of the quantum languages.
Finally, they claim that researchers should take advantage from the mistakes of the past and must rely on empirical validation when proposing new software engineering quantum techniques.
This work, however, does not rely on a documented and reproducible research methodology to find out the challenges, but leverages on the direct consequences of the statements present in the ``Talavera Manifesto''.

Zhao \etal \cite{zhao2020quantum} provided a complete overview of quantum software engineering, including all aspects of the quantum software life cycle (requirements analysis, design, implementation, testing) as well as the critical topic of quantum software reuse. The report also explored some of the challenges and opportunities of the field.
They recognize as critical that a comprehensive software engineering discipline emerges for the development of quantum software.
\revtwo{Moreover, this work defines the life cycle of quantum software and leverages it as a basis for a comprehensive survey of current research efforts in quantum software engineering. 
It also overviews quantum software testing and maintenance and discusses some fundamental issues in quantum software engineering.}
Similarly to Piattini \etal's work \cite{piattini2021toward}, this study does not exploit an empirical study to discover the challenges of the field, but relies on the open questions posed by the surveyed studies.

The first attempt to understand the challenges of quantum programming from a developer perspective was made by El Aoun \etal \cite{Elaoun2021icsme}.
The paper reports on an empirical investigation conducted analyzing the Stack Exchange forum, where developers ask and answer QSE-related topics, and the \textsc{GitHub} issue reports, where developers raise QSE-related concerns in real-world quantum computing projects.
First, the authors reviewed the categories of QSE-related questions raised on Stack Exchange, based on an existing taxonomy of question types on Stack Overflow.
Afterward, the subjects of QSE-related Stack Exchange posts and GitHub bug reports were discovered using automated topic modeling.
Their main findings showed that the most asked and answered issues in online forums are related to the theory behind quantum programming, the usage of specific data structures and algorithms, the implementation of quantum-related tasks, and the lack of learning resources.
This work differs from ours from a methodological perspective: El Aoun \etal derive their set of challenges in QSE from an automated approach (\ie topic modeling), whilst in our work, the taxonomy of quantum challenges is derived from what developers pointed out as a challenge, having explicitly asked them. Under this perspective, these two approaches can be seen as complementary.

Finally, some other studies in the field of quantum software engineering have tackled a specific challenge, and proposed preliminary solutions.
For instance, some have faced the challenge of artifact modeling~\cite{barbosa2020software, exman2021quantum, perez2021modelling, gemeinhardt2021towards}, and others have discussed quality issues, ranging from the definition of specific metrics, to debugging~\cite{zhao2021identifying, zhao2021some, campos2021qbugs}.

Our work puts its foundations on the idea that quantum programming poses some challenges that research in quantum software engineering should face.
However, differently from previous studies, we conducted a preliminary software repository mining investigation, to better understand the current usage of quantum programming technologies, and then we surveyed quantum developers, asking directly to them what they perceive as a challenge in their field.

% !TeX spellcheck = en_US
\section{\revised{Research Methodology}} \label{sec:design}

The \textit{goal} of our study was to investigate the current usage of quantum programming technologies and to explore the challenges that quantum developers face nowadays, with the \textit{purpose} of assessing where software engineering methods and practices can be applied and how. The \textit{perspective} was of researchers, practitioners, and tool vendors: the former are interested in understanding how software engineering could steer the research in the quantum programming field, letting the challenges emerge; practitioners are interested in gathering insights on how to engineer software products that include quantum computing components; tool vendors are instead interested in assessing the current support provided to quantum developers, to understand possible addressable limitations. 

\begin{figure}[t]
	\includegraphics[width=\columnwidth, clip]{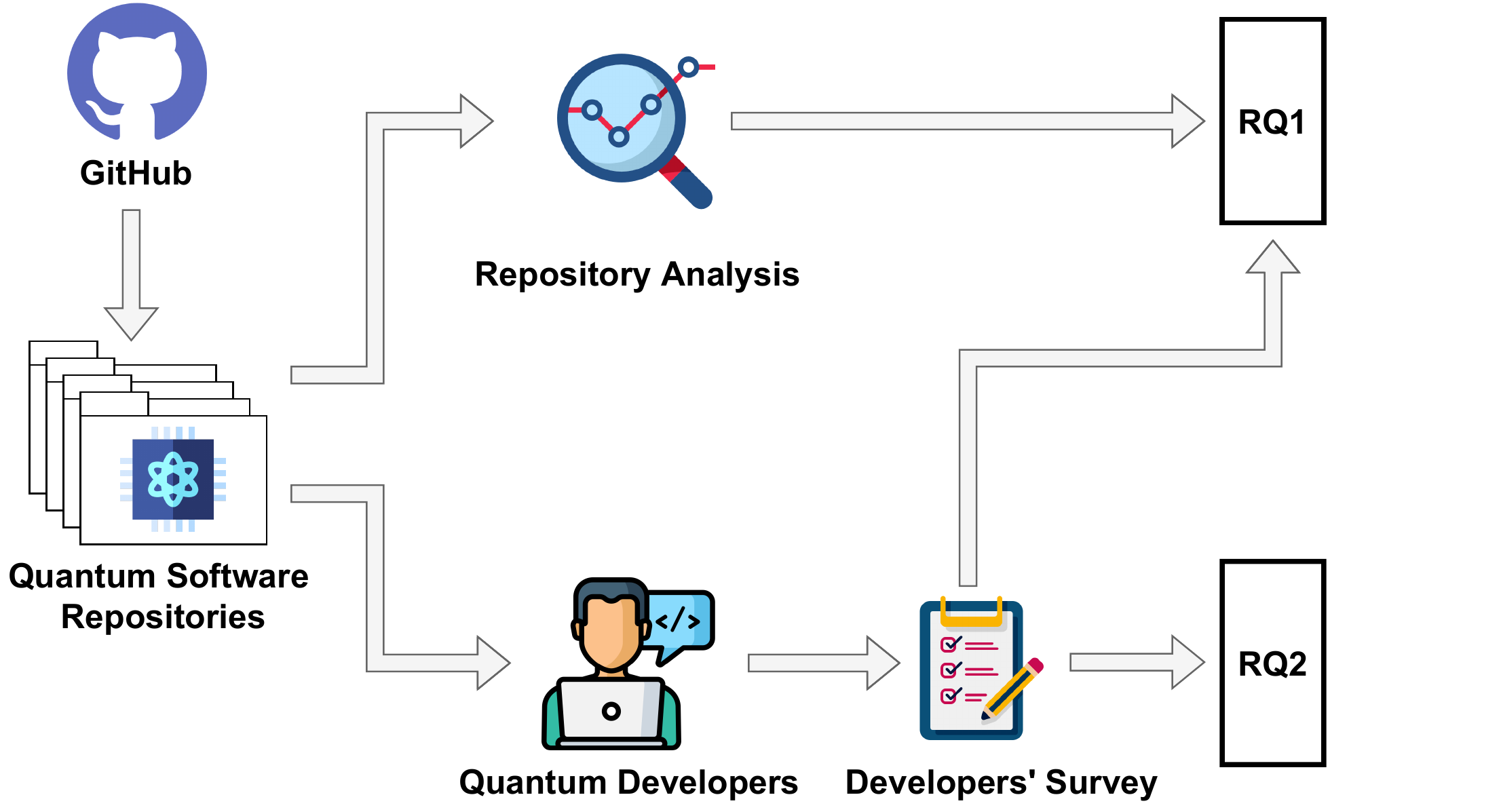}
	\caption{\revtwo{Graphical summary of the applied methodology. By mining \textsc{GitHub}, we have obtained a set of repositories that we manually classified. Afterward, we extracted the list of public emails of their developers, whom we surveyed. We addressed RQ$_1$ using the classification results, while we addressed RQ$_2$ with the survey responses.}}
	\label{fig:methodology}
\end{figure}

More specifically, our work was structured around two main research questions. \revised{We started our investigation by assessing the current usage of quantum programming, namely, the developer's purpose to employ quantum computing frameworks in their products. An improved understanding of these aspects may provide insights into the activities developers typically perform with the support of quantum computing and where software engineering might be more needed. Specifically, we asked:}

\smallskip
{\centering
\begin{rqbox}
    \textbf{RQ$_1$.} \emph{To what extent and for what \revised{purposes} are quantum programming frameworks being used?}
\end{rqbox}
}

Once we had investigated the current adoption of quantum programming in practice, we deepen our analysis to identify the major challenges that quantum developers face when dealing with the frameworks that make quantum computing accessible and usable. Knowing the challenges that developers typically encounter would help tool vendors better assist developers and researchers in devising novel techniques and approaches to deal with those issues. Hence, we formulated the second research question: 

\smallskip
{\centering
\begin{rqbox}
    \textbf{RQ$_2$.} \emph{What are the main challenges that quantum developers are experiencing when interacting with quantum frameworks?}
\end{rqbox}
}

We aimed at elaborating on the key challenges and opportunities for software engineering when it turns to quantum programming. In particular, we sought to elicit a comprehensive set of challenges that would make quantum developers' daily programming tasks easier if addressed.

As an ultimate result of our empirical investigation, we provided an improved view of how quantum programming currently is and how it might potentially be, should a software engineering process be successfully adopted. We approached such an objective using multiple research instruments, as depicted in \Cref{fig:methodology} and detailed in the next sections. In a nutshell, we employed software repository mining and grounded theory to classify the current adoption of quantum computing frameworks; then, we set up a survey study to elicit the software engineering challenges that quantum developers face.
By design, our empirical study can be classified as a mixed-method investigation where both quantitative and qualitative methodologies are employed~\cite{johnson2004mixed}. In terms of reporting, we followed the recent ACM/SIGSOFT Empirical Research Standards~\cite{ralph2020acm}.\footnote{Given our study and currently available standards, we followed the general guidelines when reporting the study design and results.}

\subsection{Context of the Study}
\revised{As explained in \Cref{sec:background}, universal gate quantum programming is based on APIs provided by third-party libraries. For this reason, the scope of the work is mainly delimited by the quantum technologies considered. To understand the current adoption and challenges of quantum programming, we focused on understanding how third-party libraries are used and their limitations from both a technological and a socio-technical perspective. Hence, the \emph{context} of the study was focused on the three state-of-the-practice universal gate quantum programming technologies, \ie \qiskit~\cite{aleksandrowicz2019qiskit}, \cirq~\cite{cirq_developers_2021_4750446}, and \qsharp~\cite{microsoft_quantum}. These three frameworks represent the main instruments that quantum developers can currently use. As shown later in the paper, they are indeed used by over 95\% of the open-source projects on \textsc{GitHub} that include quantum components.}

The frameworks are developed and maintained by three big corporations that are widely involved in quantum computing technologies, namely \textsc{IBM}, \textsc{Google}, and \textsc{Microsoft}. These technologies are widely recognized as more mature and stable than others~\cite{quantum1,quantum2}, having their own peculiar functionalities, and allowing to write and run quantum programs on both local simulators and real quantum devices provided by their vendors.

\subsection{\revised{Data Collection}} \label{sec:collection}
\revised{In our empirical study, we exploited two main sources of information, \ie the \emph{quantum software repositories} and the \emph{quantum developers}.
In the following, we describe the steps we performed in the data collection phase, whilst the data analysis process is described in the next section.}

\subsubsection{\revised{Quantum Software Repositories}} \label{sec:dataAnalysisMiningPart}
To understand the extent to which and how quantum frameworks are currently used in practice, we first adopted a software repository mining approach aiming at finding those projects which are hosted on \textsc{GitHub} that use at least one of the considered technologies.

By means of the \textsc{Python} Client Library of the \textsc{GitHub} REST APIs,\footnote{\texttt{PyGitHub}: \url{https://github.com/PyGithub/PyGithub}} we searched for code snippets (using the \texttt{search\_code} function) that indicated the use of the technologies we were interested in. More specifically, for \qiskit and \cirq we looked for patterns like \texttt{`from qiskit import'} and \texttt{`from cirq import'}, respectively. Both frameworks are written in \textsc{Python} and, therefore, developers must necessarily use those patterns to include the libraries in their programs. In other words, the use of these patterns ensured the identification of all the repositories that currently employ \qiskit and \cirq in their code. When it turns to \qsharp, we had to use a different strategy. Unlike the others, \qsharp is recognized by \textsc{GitHub} as a programming language. For this reason, we could directly look for the repositories written in such a language: we used the \texttt{search\_repositories} function, passing \qsharp as language parameter. In our online appendix~\cite{appendix}, we released the source code developed for mining quantum repositories.
Overall, this process found a total of \totalrepos unique repositories (442 using \qiskit, 217 using \cirq, 72 using \qsharp).

\subsection{\revised{Quantum Software Developers}} \label{sec:surveyDesign}

\begin{table*}[!th]
	\centering
	\caption{Questions asked in the survey} \label{tab:survey}
	\resizebox{\textwidth}{!}{
		\begin{tabular}{@{}p{0.6\textwidth}lp{0.3\textwidth}@{}}
			\toprule
			\bfseries Question Text & \bfseries Answer Type & \bfseries Possible Answers\\
			\midrule
			\multicolumn{3}{@{}l@{}}{\textit{\bfseries Part 1 - Background}}\\
			What is your current employment status? & Multiple Choice & B.Sc. Student; M.Sc. Student; Ph.D. Student; Researcher; Open Source Developer; Industrial Developer; Other          
			\\ \addlinespace[0.5em]         
			What is your educational background? & Single Choice & Computer Science; Chemistry; Physics; Other
			\\ \addlinespace[0.5em]
			What is your age range? & Single Choice & 18-24; 25-34; 35-44; 45-54; 55+
			\\ \addlinespace[0.5em]
			What is your gender? & Free Text & -
			\\ \addlinespace[0.5em]
			Please, indicate your expertise (in years) in Software Development. & Single Choice & None; 0-3; 3-5; 5-10; 10+
			\\ \addlinespace[0.5em]
			Please, indicate your expertise (in years) in Industrial Development. & Single Choice & None; 0-3; 3-5; 5-10; 10+
			\\ \addlinespace[0.5em]
			Please, indicate your expertise (in years) in Quantum Programming. & Single Choice & None; 0-3; 3-5; 5-10; 10+
			\\ \addlinespace[0.5em]
			What is your Country? & Free Text & -
			\\ \addlinespace[1em]
			\midrule
			
			\multicolumn{3}{@{}l@{}}{\textit{\bfseries Part 2 - Current Adoption}}\\
			Which quantum technology are you most confident with? & Single Choice & \qiskit; \cirq; \qsharp; Other
			\\ \addlinespace[0.5em]
			Which other quantum technology do you use? & Multiple Choice & \qiskit; \cirq; \qsharp; Other
			\\ \addlinespace[0.5em]
			In which context are you using quantum computing? & Multiple Choice & Academic Study; Hackaton; Industry; OSS; Personal Study; Research; Other
			\\ \addlinespace[0.5em]
			Could you please tell more about the tasks you are performing with quantum computing? & Long Free Text & -
			\\ \addlinespace[1em]
			\midrule
			\multicolumn{3}{@{}l@{}}{\textit{\bfseries Part 3 - Potential Adoption and Challenges}}\\
			Consider the technology you are most confident with. What were the top 3 challenges that you have faced? & Multiple Free Text & -
			\\ \addlinespace[0.5em]
			Based on your experience, have you ever solved (or tried to solve) a problem using quantum programming which has no "traditional" solution (or the solution is intractable)? & Single Choice & Yes; No
			\\ \addlinespace[0.5em]
			If yes, could you please elaborate on the problem and why you have to use quantum computing? & Long Free Text & -
			\\ \addlinespace[0.5em]
			Based on your experience, have you ever solved (or tried to solve) a problem that has a "traditional" solution using quantum programming? & Single Choice & Yes; No
			\\ \addlinespace[0.5em]
			If yes, could you please elaborate on what it was and explain why you chose to use quantum computing? & Long Free Text & -
			\\ \addlinespace[0.5em]
			
			\bottomrule
		\end{tabular}
	}
\end{table*}

In the context of our empirical investigation, we needed to collect opinions from developers of quantum-based applications. This step was harder than expected since quantum programmers are a new category of developers that are not necessarily computer scientists but also physicians, chemists, and others. Therefore, we had to customize the recruitment of the participants, the design of the survey, and its dissemination.

\smallskip
\textbf{Survey Recruitment.} We took advantage of the software repositories mined from \textsc{GitHub} to obtain a list of eligible candidates for our survey.
In this way, we ensured to involve developers having some real experience with quantum programming.
Starting from the initial set of 2,399 unique developers, we selected only developers having a public email to be contacted with.
The email address collection was done by exploiting the \textsc{GitHub} APIs.
By applying this selection criterion, we obtained a set of 984 developers. Of these, \textcolor{black}{79} had less than ten commits: we excluded these developers as they might have only tangentially contributed to the projects and, as such, they might not have the adequate expertise and/or experience to address our questions~\cite{sugar2014studies}.
This filter led to a total of \textcolor{black}{905} contributors which could participate to our survey.

We employed an \emph{opt-in} strategy~\cite{hunt2013participant} when involving developers. We sent a first email asking whether they would have liked to participate in our survey study and, only in case of positive feedback, we sent them additional instructions.
%\revised{This was an important selection criterion: we could not explicitly conatct and ask to participate to our survey to people without the email.}
With this strategy we recruited 56 \emph{volunteers} and mitigated possible legal concerns~\cite{surveyIssues}. Nonetheless, it might have led to self-selection or voluntary response bias~\cite{heckman1990selection,sakshaug2016evaluating}. To mitigate this, we introduced a prize of four Amazon gift cards with a total value of \$100.

\smallskip
\textbf{Survey Design.} The survey was composed of three main sections, as reported in \Cref{tab:survey}.
The first one aimed at gathering the background of the participants. Besides asking about their current employment and their experience with software development, industrial development, and quantum programming, we asked for information about gender and country (in a free text form, as recommended by recent research~\cite{fink2003survey,broussard2018too}) and educational background. This latter question was intended to understand more closely the expertise of the current population of quantum programmers; such information might be interesting to reveal cultural barriers to quantum software engineering~\cite{noll2011global}. 

The second section of the survey aimed at gathering developer's opinions about the current use of quantum technologies. We first asked which of the available frameworks they use and, more in general, the technologies they are typically comfortable with. Afterward, we asked the reason behind the use of quantum programming. This question was essential to understand better why quantum technologies are currently employed, \eg for an academic interest or for a personal study of how to program using quantum computing. In addition, this question enabled us to complement the grounded-theory exercise conducted to address \textbf{RQ$_1$} (more details are reported in \Cref{sec:dataAnalysis}).

The last section targeted the longer-term adoption of quantum technologies. More particularly, it aimed at assessing the main challenges currently faced by developers when adopting the quantum framework they are most comfortable with. This question sought to elicit the most relevant limitations of the existing instruments, potentially highlighting the most pressing issues and challenges that researchers and tool vendors might need to pay attention to. Afterward, we focused on the interplay between quantum and traditional computing. Finally, it aimed at analyzing whether developers addressed or are trying to address problems with and without traditional or tractable solutions through the adoption of quantum technologies. In both cases, we asked to further elaborate on the specific problems treated and the rationale behind the use of quantum programming. Answers to these questions helped us to understand how far quantum technologies are from solving real-case problems and the software engineering methodologies and tools required to support developers. 

\smallskip
\textbf{Survey Dissemination.} The survey was developed and disseminated to participants using \textsc{LimeSurvey},\footnote{\textsc{LimeSurvey} website: \url{https://www.limesurvey.org}.} an open-source survey editor released under GNU-GPL license. The questionnaire was available from June 1 to June 30, 2021. The survey link was sent to every recruited developer via e-mail. We estimated a completion time of 15 minutes.

\smallskip
\textbf{Ethical and Privacy Considerations.} In our country, it is not mandatory yet to seek approval from an Ethical Review Board when releasing surveys and experiments with human subjects. Nonetheless, we took into account several possible ethical and privacy concerns. In the first place, we guaranteed the participants' privacy by not asking their names or email addresses, hence gathering anonymous answers. When recruiting developers, we clearly stated the goal of the survey study, as well as explicitly report that the given answers would have been used in the scope of a research activity that would not have any intention of publishing sensitive data. We also clarified that completed surveys would eventually become public, although guaranteeing their privacy. Finally, opting for an open-source survey tool avoided ethical concerns that might have potentially led participants to feel uncomfortable answering questions using commercial editors (\eg \textsc{Google Forms}). Indeed, previous work~\cite{buchanan2009online} showed that this aspect impacts the potential response rate in survey studies.

\subsection{\revised{Data Analysis}}
\label{sec:dataAnalysis}
\revised{Once we had collected all the required data, we addressed our research questions. To answer the first research question (\textbf{RQ$_1$}), we employed the information coming from both the repository mining and the first two parts of the developers' survey. To answer the second research question (\textbf{RQ$_2$}) we only relied on the responses provided to the third part of the survey.}
\revised{The investigation into the purposes of the repositories collected from \textsc{GitHub} and the challenges from the developer's survey share the same methodology. In particular, in both cases, we applied a systematic approach that constructs theories applying a methodical gathering and analysis of the data, known as Straussian Grounded Theory~\cite{corbin1990grounded}. This methodology does not assume previous theoretical formulation but rather adopts an approach wherefore a theory is directly and solely generated from data. As for the other data collected from the survey, \ie the background and current adoption of quantum programming, we mainly relied on statistics. More details are reported in the following.}

\begin{description}[leftmargin=0.3cm]
	\item[\revised{Application of Straussian Grounded Theory.}] \revised{The first author (\ie hereafter, \emph{the main inspector}) extracted the excerpts from the textual data we had.}
	\revised{In the context of the repository analysis, the excerpts were \texttt{README} files and repository descriptions (translated in English if written in another language) of each analyzed repository.
	If both \texttt{README} file and the repository description were missing, we labeled the purpose of the repository as ``Unknown''. As for the survey, the excerpts were the individual answers provided by developers.} 
	\revised{The main inspector applied open coding \cite{corbin1990grounded}, which is the process of taking the excerpts and continuously comparing and contrasting them with other excerpts, with the final aim of grouping them, and thus giving them a code.
    These excerpts were considered based on semantic similarity~\cite{harispe2015semantic}. This process was also carried out individually by the other two inspectors (the second and third author of the paper) to have coding as unbiased as possible.
    To validate this step, we also computed their agreement in terms of Fleiss' \emph{k} \cite{fleiss1971measuring}.}
	
	\revised{Once the inspectors assigned the codes to the excerpts, they proceeded with the axial coding step \cite{corbin1990grounded}, which consists of grouping into categories the codes found in the previous step. 
    Alongside the open coding, it is a cyclical process since new excerpts and examined categories might contradict, support, or expand the existing codes and categories. When additional excepts do not expand upon the found code and categories, theoretical saturation~\cite{corbin1990grounded,walker2012research} is reached, \ie the point at which codes and categories are stabilized and fixed.}
	
	\revised{The resulting taxonomy collects all the codes and categories, resulting from the open and axial coding steps, linked together by a common aspect (or core category), found by selective coding \cite{corbin1990grounded}.
    This core category represents the foundation on which the answers to our research questions rely.}

	\item[\revised{Analysis of Other Survey Information.}] \revised{We analyzed the first two sections of the survey to understand the background of the developers in our sample and their take on the current adoption of quantum programming. The demographics of the surveyed developers and their current employment status were summarized using statistics. Furthermore, we analyzed the distribution of contributors per repository category and the distribution of repositories per developer. These distributions were compared against the answers obtained from the survey to add qualitative insights to the repository analysis previously conducted.}
	
\end{description}	

% !TeX spellcheck = en_US
\section{\revised{RQ$_1$. On the Current Usage of Quantum Programming}}
\label{sec:rq1}

\begin{table*}[t]
	\centering
	\caption{Summary of the labels employed in the classification of the mined repositories} \label{tab:repolabels}
	\resizebox{\textwidth}{!}{
		\begin{tabular}{@{}lp{0.4\textwidth}p{0.4\textwidth}@{}}
			\toprule
			\bfseries Label Name & \bfseries Purpose & \revised{\bfseries Example}\\
			\midrule
			Exercise/Toy & Repository containing toy projects or collection of sample code. & \revised{\textbf{ryuNagai/QML.} Repository containing experimental code on Quantum Machine Learning.} \\ \addlinespace[0.5em] 
			Hackaton/Assignment & Repository containing code developed for a hackaton or a school assignment. & \revised{\textbf{oliverfunk/quantum-natural-gradient} Quantum Natural Gradient implemention in Qiskit for the 2019 Qiskit Africa Camp.} \\ \addlinespace[0.5em]   
			Library/Framework & Repository containing code composing a library or a framework. & \revised{\textbf{Davidelanz/quantum-robot.} Package for quantum-like perception modeling for robotics.}\\ \addlinespace[0.5em]  
			Research & Repository containing code belonging to a paper or research appendix. & \revised{\textbf{BramDo/custom-cx-gate-on-Casablanca.} Code of Qiskit Pulse - Programming Quantum Computers Through the Cloud with Pulses.} \\ \addlinespace[0.5em]    
			Teaching & Repository containing code that complements a lecture or a textbook. & \revised{\textbf{kongju/QML.} Lecture notes and the code for the course on Quantum Machine Learning offered by University of Toronto on edX.} \\ \addlinespace[0.5em]             
			Tool & Repository containing code for a supporting tool, \eg a quantum compiler. & \revised{\textbf{mtreinish/bqskit-qiskit-synthesis-plugin.} This repository contains a PoC unitary synthesis plugin for Qiskit.} \\ \addlinespace[0.5em]    
			Unknown & Repository containing code not classifiable by reading the README or the Description \revised{or not having any of them.} & \revised{\textbf{eggerdj/backends.} Repository missing both the README and the Description.} \\ \addlinespace[0.5em]   
			\bottomrule
		\end{tabular}
	}
\end{table*}

\begin{figure}[t]
	\centering
	\resizebox{0.43\textwidth}{!}{
		\includegraphics{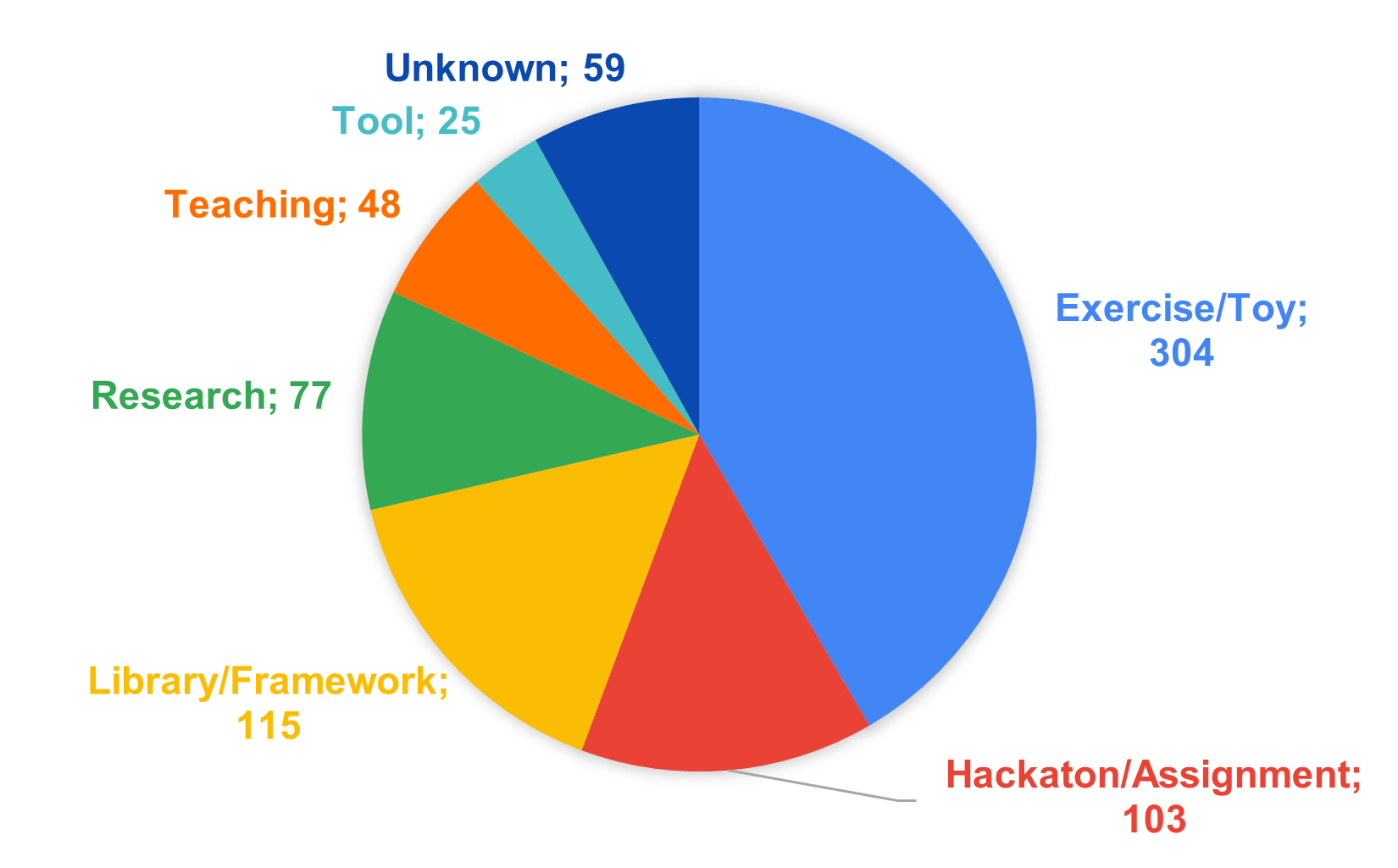}
	}
	\caption{Partition of the repository per class of usage.} \label{fig:task1}
\end{figure}

\begin{figure}[t]
	\centering
	\resizebox{0.92\columnwidth}{!}{
		\includegraphics{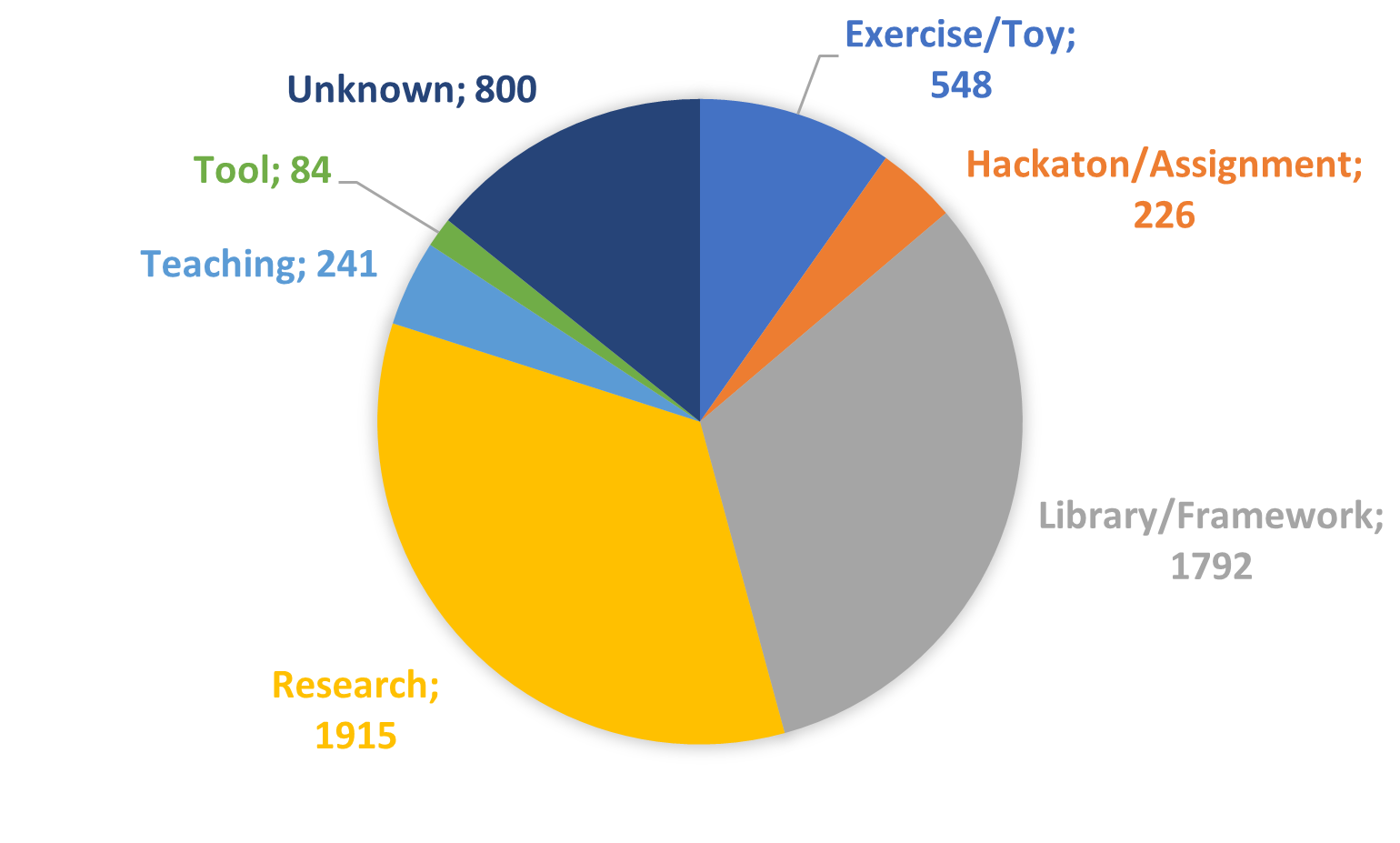}
	}
	\caption{Number of contributors per type of repository.} \label{fig:devs_per_class}
\end{figure}

\begin{figure}[t]
	\centering
	\resizebox{0.9\columnwidth}{!}{
		\includegraphics{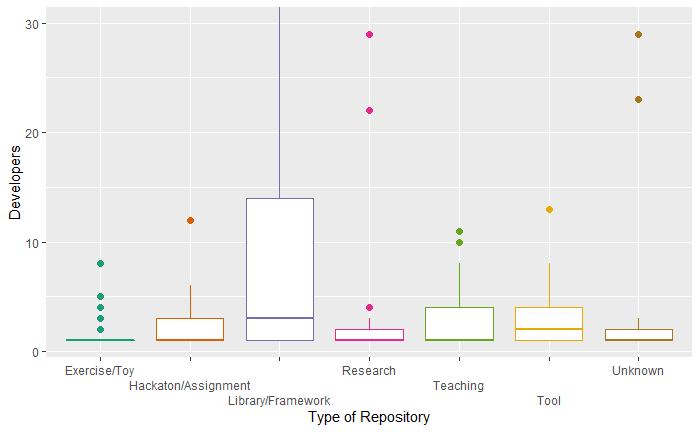}
	}
	\caption{Distribution of developers by kind of repository. Toy projects have a distribution skewed on the minimum (1), whilst Framework ones have a much wider distribution of contributors.} \label{fig:devs_distribution}
\end{figure}

\begin{figure}[t]
	\centering
	\resizebox{0.9\columnwidth}{!}{
		\includegraphics{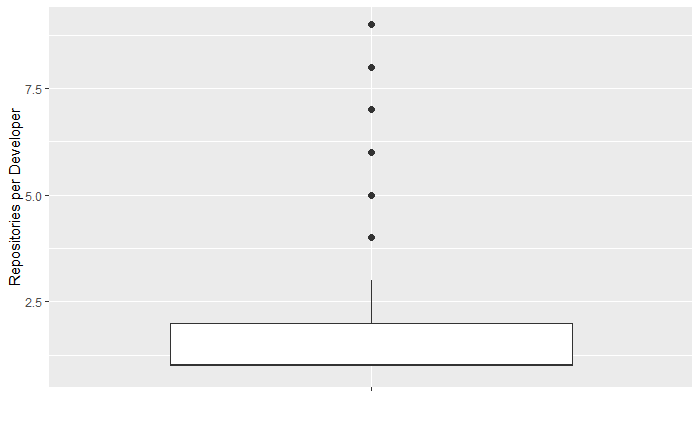}
	}
	\caption{\revised{Distribution of repositories per single developer. The median value insists on 1, indicating that most quantum developers have contributed only to a single repository.}} \label{fig:repo_per_dev}
\end{figure}

The analysis of \totalrepos repositories led to a taxonomy of the current usage of quantum programming technologies, reported in \Cref{tab:repolabels}.
The taxonomy is composed of six categories, representing the high-level purpose for which the repository was created, \eg study purposes.
\Cref{fig:task1} summarizes the repositories partitioned using the taxonomy that we have built, \Cref{fig:devs_per_class} plots the distribution of developers per kind of repository, whilst \Cref{fig:devs_distribution} depicts the distribution of developers by type of repository.
It must be pointed out that the total number of unique contributors found is 2,399, which is not obtainable by summing up all the values reported in \Cref{fig:devs_per_class}.
The reason is that a developer might be contributing to more than a single repository.
To this extent, we also analyzed the number of repositories for single developers, whose distribution is reported in \Cref{fig:repo_per_dev}.
\revised{It is possible to note that, although some developers might have been counted more than once in \Cref*{fig:devs_per_class}, the majority we took into consideration has contributed just to a single repository. Thus the number of developers per class of repositories is much more significant.}

Since quantum programming is still in its infancy, the main purpose of use is for exercise or personal study.
The hosted code aims to explore the features of quantum programming and, in general, is not intended to become a real-world software. 
This is proved by the fact that 41\% of the analyzed repositories belong to this category. 
However, as shown in \Cref{fig:devs_per_class} only 548 developers contribute to this kind of repositories, since most of these repositories have only one single contributor, as shown in \Cref{fig:devs_distribution}.
Some example of projects developed for such purpose are quantum games, \ie small and generally text-based games, which leverage the power of \textit{real} randomness given by the qubits to generate random worlds, or scenarios.
These are directly correlated with the third-largest slice of repositories, which is composed of repositories developed during hackathons or class assignments.

The other main purpose for which quantum programming is used is to develop quantum libraries or frameworks, which represent the 16\% of the total repositories.
This outcome is indeed reasonably expected since quantum technologies currently under development are mostly open-source (\eg \qiskit, \cirq).
Moreover, domain- and task-specific libraries are also emerging (\eg quantum machine learning or chemistry ones).
This is the second top class of repositories for number of contributors (1792), and the class of repositories having the most variable distribution of contributors, with a median value of 3.

Online appendices of research papers or related research projects represent 11\% of the considered repositories.
This result is in line with the fact that quantum programming is still a neat field in the vast plethora of computer science and physics research.
Although being the top class for number of developers (1915), the distribution of developers ranges from 1 to 375, with 1 as median value.

The remaining cases of use of quantum programming represent only a small percentage over the total.
Evidence has been observed of material used for teaching purposes: book appendices, blog posts, etc. represent 7\% of the analyzed repositories.
There are also some tools among the repositories that we have inspected (3\%), but these, along with teaching materials, were never mentioned by the surveyed developers.

\begin{figure}[t]
	\centering
	\resizebox{0.72\columnwidth}{!}{
		\includegraphics{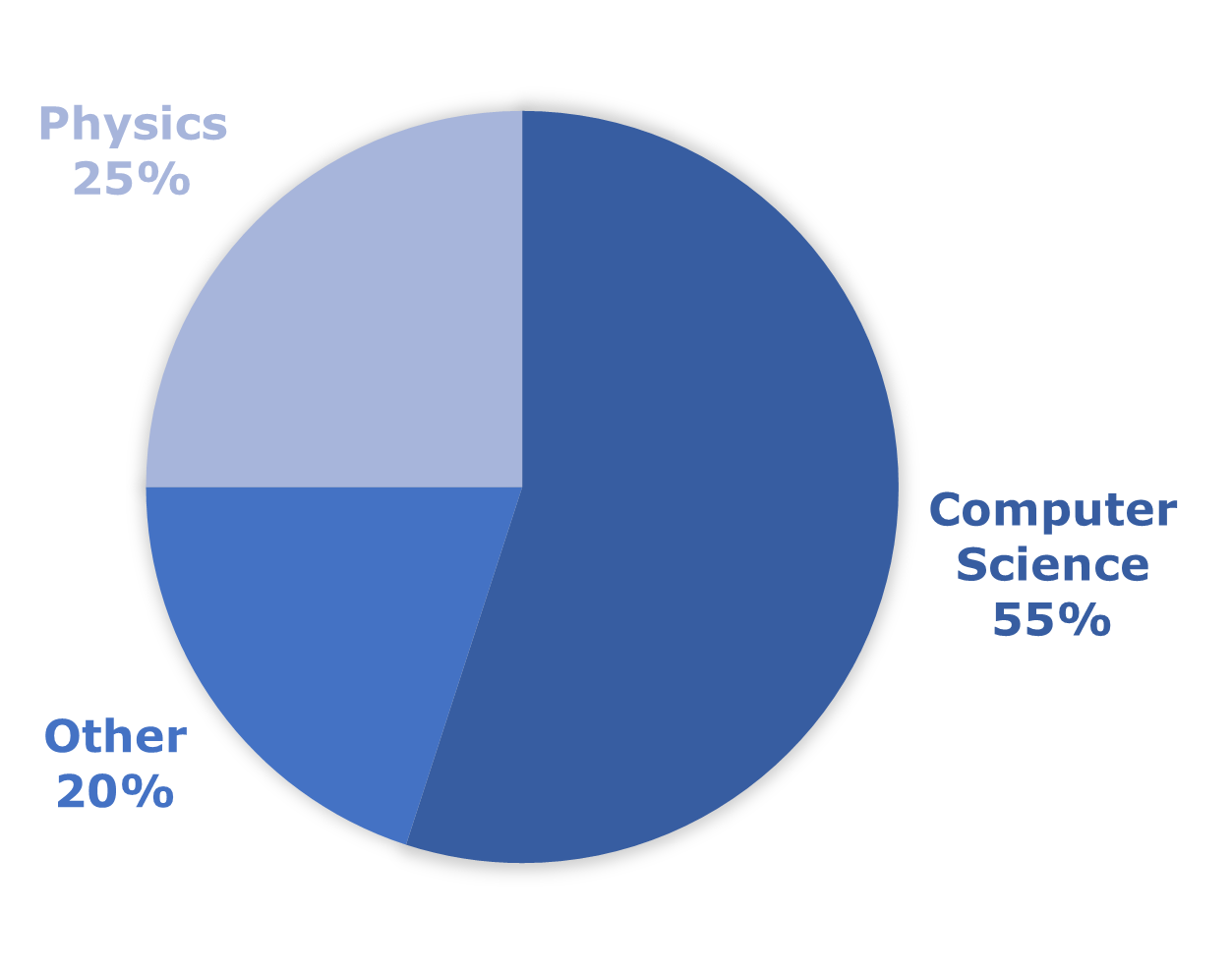}
	}
	\caption{Educational Background of the Survey Participants.} \label{fig:background1}
\end{figure}

\smallskip
\textbf{\revised{Insights from the developers}.}
\Cref{fig:background1} depicts the educational background of the survey participants.
Most of the respondents to the survey had a computer science background (55\% of the total), while 20\% had a physics one.
The remaining 25\% of the respondents have other backgrounds, \eg Mathematics, electrical engineering, civil engineering.
71\% of the participants were younger than 35 years, with most of them (39\%) belonging to the 25-34 age range.
Another 21\% have an age between 35 and 44 years.
Finally, 4\% of them belong to the 45-54 age range; the last 4\% are older than 54 years.

\begin{figure}[t]
	\resizebox{\columnwidth}{!}{
		\includegraphics{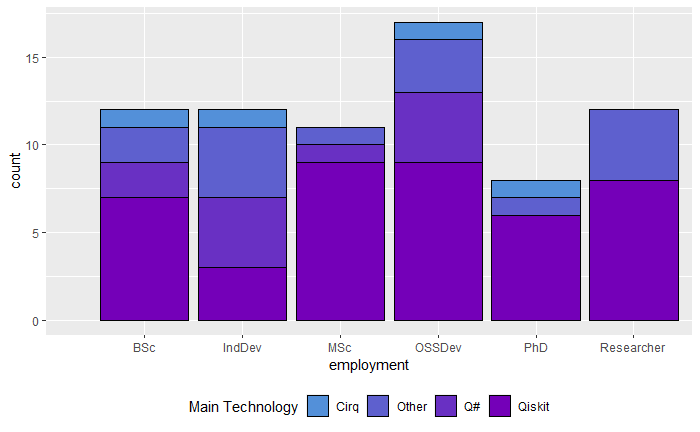}
	}
	\caption{Employment Status of the Survey Participants, \revised{divided by main tchnology used.}} \label{fig:background2}
\end{figure}

\Cref{fig:background2} shows the employment status of the participants at the time of the survey, \revised{divided by used technologies}.
Most of the participants are Open Source Developers (17), although the majority of them (31) are involved in academia (12 BSc Students, 11 MSc Students, 8 Ph.D. Students).
\revised{Researchers and Industrial developers include 12 people each.
It is also possible to notice that the most used technology is indeed \qiskit in almost all the employments categories, except for industrial developers, who mostly use \qsharp and other technologies.
\cirq is the least used technology by all the categories.
It is worth pointing out that researchers only employ \qiskit and \qsharp.}  
Most of the respondents (27\%) come from the USA, while the other countries with the highest number of respondents are India (12\%) and Italy (9\%).
The other respondents mainly come from European countries (\ie France, Germany, Greece, Poland, Portugal, Spain, Sweden, Switzerland, United Kingdom), Eastern countries (\ie China, Japan, Israel, Turkey), and American countries (\ie Canada, Colombia, Mexico).

\begin{figure*}[t]
	\centering
	\resizebox{\textwidth}{!}{
		\includegraphics{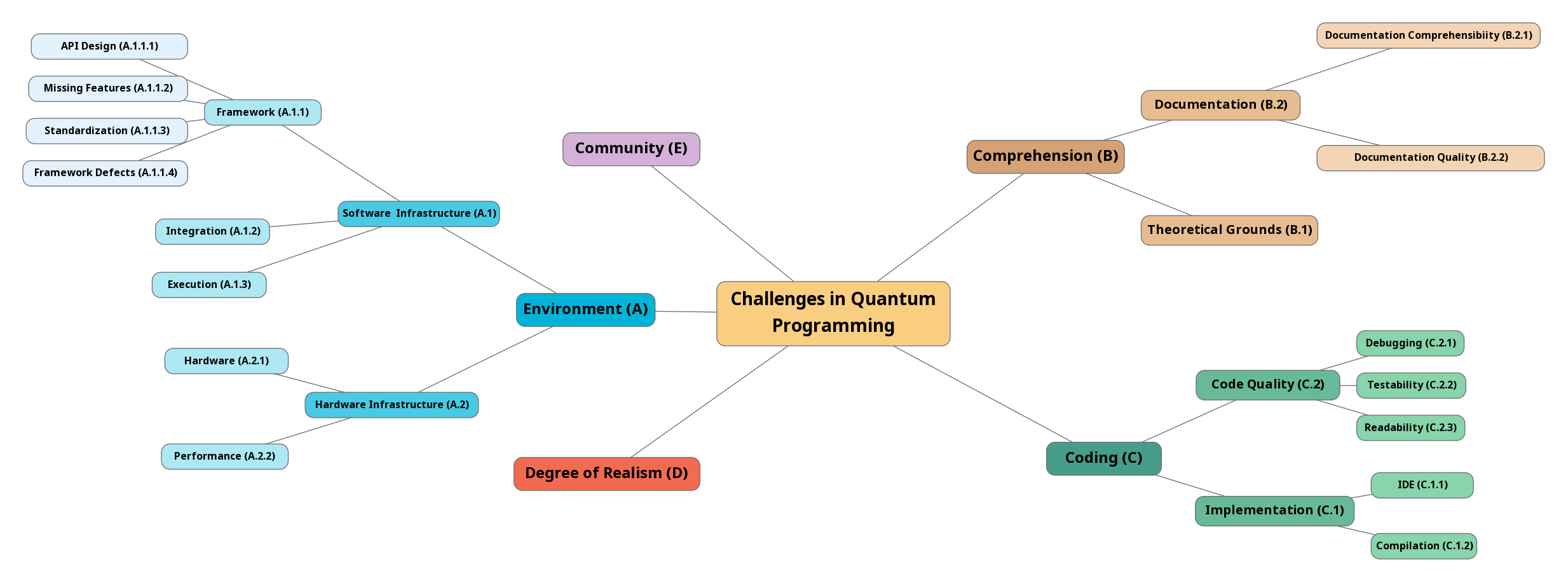}
	}
	\caption{Graphical representation of the obtained taxonomy of quantum computing challenges.} \label{fig:taxonomy}
\end{figure*}

The results of the survey show that most of the interviewees (29\%) indeed use quantum programming for personal learning.
Another large group of users of quantum programming belongs to those who carry out research, who are divided into two categories, namely those who use quantum programming \emph{for} research, and those who actually do research \emph{on} quantum programming.
The first group is represented by 14\% of the interviewees, while the second by 23\%, for a total of 37\% developers using quantum programming for research activities.
Another 14\% of survey respondents use quantum programming to develop technologies for quantum programming itself.

Another interesting result that emerges by analyzing of the answers given to the last questions of the third part of the survey, \ie whether the respondent has ever applied quantum computing for an intractable classical computation problem (and what was it), and whether the interviewee has ever applied quantum computing to a problem solvable with classical computation (and what was it).
The analysis of these answers show that few respondents have tried it, and the problems solved are actually those which are used as an example to show the advantages of quantum computation over the classical one (\eg applying the Shor algorithm for the factorization).
This reinforces the idea that the current use of quantum programming is mainly didactic.

\smallskip
\begin{summarybox}[RQ1]
	Quantum programming technologies are used mostly for personal study purposes: 41\% of the analyzed repositories were created for this.
	On the other hand, framework and research repositories have the largest number of contributors, meaning that a major effort is put in these activities.
	\revised{The conclusion is supported by the survey that showcased that most respondents were involved in research or frameworks development activities.}
\end{summarybox}

% !TeX spellcheck = en_US
\section{\revised{RQ$_2$. On the Current Challenges of Quantum Programming}}\label{sec:taxonomy}
\Cref{fig:taxonomy} depicts a hierarchical taxonomy of the challenges in quantum programming that we have constructed as outcome of the Straussian Grounded Theory excercise described in \ref{sec:design}.
Some categories have just one layer, while others have been partitioned.
In particular, some categories presented a set of challenges that were much more cohesive than others, and thus they do not need a specialized category to be described.
In the following, we describe each category (alongside their sub-categories) of challenges.

\smallskip
\noindent \faHandORight \xspace \textbf{Environment (A)}
The problems described in this macro-category are all related to the \textit{quantum} environment, both in terms of hardware and software.

\begin{description}[leftmargin=0.3cm]
	
\item[Software Infrastructure (A.1)] The challenges in this area range from those relating to the framework being utilized to those relating to the execution environment.\\	
\textbf{Framework (A.1.1).} Challenges regarding the frameworks are mainly related to API design, missing features and standardization.
In particular, 15 participants mostly complained about the `\textit{`constant changes in API}'' of quantum technologies, which often are not correlated with an adequate ``\textit{deprecation policy}''.
Other 2 developers complained of lacking features.
For instance, ``\textit{some kind of operations are not directly supported by \qiskit}'', or ``\textit{Are not supported quantum operations among different circuits, which could be interesting to develop operations in different nodes, for Quantum Networks}''.
A set of 12 participants also pointed out challenges concerning ``\textit{gate definitions}'' and ``\textit{endianess of the qubits}'', which differ from framework to framework. \ie standardization problems.
Some technologies implement concepts (\eg quantum gates) differently from how they are defined in theory and thus hindering the actual development process.
``\textit{Conventions and standards haven't fully solidified across the industry yet}''.
Standardization challenges are revealed also in the form of lack of abstraction.
At the time of writing, quantum programs are generally written in quantum circuits interacting with classical code.
To some extent, a quantum computer is programmed at a very low level of abstraction compared to traditional computers.
Having a low level of programming inevitably raises some issues, such as code portability and ``interfacing with a wide variety of hardware platforms''.
In general, as other developers pointed out, quantum code written with a specific technology is meant to be run on the machines that the technology vendor offers, thus leading to \textit{vendor lock-in}.
Finally, although some technologies offer ready-to-use implementations, developers must implement quantum algorithms from scratch, relying on what the platform offers.
Some developers, finally, (4) complained about defects within the used framework (\eg ``\textit{Incorrect noise modeling of CX gates in parallel}'').
\\
\textbf{Integration (A.1.2) .} All challenges related to integrating quantum systems with traditional ones were reported by 6 participants. 
Developers, for instance, found it challenging to ``\textit{integrating a classical algorithm into its quantum analog}'' or  `` \textit{Connecting quantum computers to blockchain networks}''.
\\
\textbf{Execution (A.1.3).} The reported challenges falling in this category are related to the execution environment, and were reported by 11 survey participants.
They struggle with setting up execution environments, simulators, or classical systems with which quantum programs interact, hindering their ability to execute their programs.

\item[Hardware Infrastructure (A.2)]
This category groups together challenges that somehow relate to hardware aspects, ranging from the ``bare metal'' to the performances.\\
\textbf{Hardware (A.2.1)} Reported hardware challenges fall into this category.
Quantum programming requires specialized hardware, which is under constant development.
At the time of writing, we have access to quantum computers with a limited small number of qubits. Therefore, large-scale applications cannot be effectively developed.
``\textit{No real quantum computer to work on}'', ``\textit{lack of hardware support}'', ``\textit{access that is limited to small quantum devices}'' are examples of this category of challenges developers face.
Finally, quantum computers suffer from noise because of the relative \textit{immaturity} of quantum technologies and some physical limitations (\eg ``IBM’s hardware has an error rate that is too high for many purposes'', ``\textit{qubits propagate error to others while entangling}'').\\
\textbf{Performance (A.2.2).} Quantum hardware is limited, so vendors offer emulators to run quantum programs on classical computers.
This leads to another set of challenges related to performance.
Emulators are resource greedy, and limit the actual execution of quantum programs (`\textit{`Performance of emulators - even with few qubits}'', ``\textit{\qiskit objects use memory inefficiently, which made them too slow and memory hungry.}'').
Similarly, running programs on real quantum devices ``\textit{takes time to complete execution}'', since vendors queue jobs to guarantee free access to everyone.
Finally, a lack of optimization given by the vendors causes troubles to developers, who complain \textit{``Poor optimization of operations}'' and compile-time issues (``\textit{\qiskit does not have support in its pipeline for optimizing synthesis for more than 2 qubits}'').

\end{description}

\smallskip
\noindent \faHandORight \xspace \textbf{Comprehension (B)}
This macro-category refers to all the challenges that developers face in comprehending quantum programs, in all facets.
Thus, this category ranges from challenges which are inherent to the documentations of the frameworks, to challenges that are inherent to (lack of) theoretical grounds. 

\begin{description}[leftmargin=0.3cm]

\item[Theoretical Grounds (B.1)] This category includes all those challenges that are related to learning quantum technologies.
A set of 20 respondents reported that many concepts are needed to get acquainted with quantum programming, particularly linear algebra.
Moreover, it was pointed out that programming a quantum computer is completely different, and many found it challenging to theoretically understand new concepts related to quantum circuits and quantum gates.
For instance, as one developer reported, it is fairly challenging ``\textit{designing a circuit in terms of the basis gates of the QPU to reduce the computational cost to perform a desired task}''.

\item[Documentation (B.2)]
The category represents challenges related to documentation that developers face when approaching quantum technologies.
Most of the reported challenges complain about poor consistency in the documentation, and lack of proper getting started material.\\
\textbf{Documentation Comprehensibility (B.2.1).} Challenges falling into this category include inconsistent tutorials that make difficult to completely understand the technology, and, as a consequence, hinder the actual learning process of the quantum technology.
This was pointed out by 3 participants\\
\textbf{Documentation Quality (B.2.1).} Among all the participants, 16 of them pointed out problems related to the quality of the provided documentation. ``\textit{Documentation out-of-date or not comprehensive enough}'', ``\textit{Outdated and erroneous documentation}'', or ``\textit{Missing Documentation}'' are some examples of the reported complaints.

\end{description}

\smallskip
\noindent \faHandORight \xspace \textbf{Coding (C)}
In this category fall all the reported challenges related to coding quantum programs. 
\begin{description}[leftmargin=0.3cm]
	
	\item[Implementation (C.1)] Challenges in this category relate to the implementation activities.\\
	\textbf{IDE (C.1.1).} Although not necessary, it is known that a good IDE makes the difference when programming: quantum programming makes no exception.
	However, some developers (3) have pointed out some IDE-related challenges that hinder, rather than helping, their development tasks.
	For instance, many found it difficult to work with the \qsharp environment, particularly with the IDE.
	Others pointed out that working with \qiskit IDEs and Language extensions resulted in an unstable experience on particular hardware architectures.
	Finally, when it comes to plotting the defined circuit, the IDE does not allow a clear visualization, making coding even harder.\\
	\textbf{Compilation (C.1.2).} Quantum circuits are defined based on quantum technologies that must be compiled to be executed on real quantum machines.
	The current models for quantum computing require quantum algorithms to be specified as quantum circuits on ideal hardware, ignoring hardware-specific details.
	Although such modeling should make programs more portable and allow developers not to concentrate on hardware-specific issues~\cite{qcc},  such programs need to be translated into code that quantum computers can execute, \ie need to be compiled~\cite{qcc}.
	However, as 4 participants said, ``\textit{adapting ideal quantum circuits to available device architectures}'' is a tough challenge for developers, particularly, ``\textit{hard to write compiler passes}'' since ``\textit{non standard benchmarking on compilation depths}'' are available.
	
	\item[Code Quality (C.2)] The challenges that fall into this category are related to \textit{traditional} code quality problems, \eg testability, debugging, and readability.\\
	\textbf{Debugging (C.2.1) }
	Some participants (11) have pointed out that they often found it challenging to understand the error messages given by the execution of quantum programs or even the code itself. Debugging these errors is even more complicated, if we consider the uniqueness of this new programming paradigm.\\
	\textbf{Testability (C.2.2)}
	Understanding if a program performs as intended is a fundamental concept in programming.
	However, it is not so simple for quantum programming: 4 respondents have pointed out that ``\textit{being able to check that the circuit does what you want it to do}'' is particularly challenging.
	In particular, they pointed out that a major problem ``\textit{the little understanding of what should the result look like resulting in a lack of common sense regarding the correctness of the result}''.\\
	\textbf{Readability (C.2.3)}
	Another quality-related set of challenges is readability, as pointed out by 1 respondent.
	Since the quantum code consists of defining a register of qubits and applying gates on them, one of the challenges pointed out by developers is ``\textit{creating a readable code}''.

\end{description}

\smallskip
\noindent \faHandORight \xspace \textbf{Degree of Realism (D)} 
This category of challenges involves the applicability of quantum programming to solve real-world problems.
In theory, quantum computers should allow developers to solve problems that classical computers cannot solve. 
Nevertheless, in practice, there are still hardware and performance limitations, which were pointed out by 10 respondents.
``\textit{Finding interesting use cases}'', ``\textit{Using quantum computers on real products}'', or ``\textit{Formulate a problem}'' are only a few examples of the kind of challenges reported by developers.
Many people also find it challenging to design quantum programs able to solve real problems or design their quantum algorithms since ``\textit{There aren't yet many problems that quantum can solve that traditional tech can't}''.

\smallskip
\noindent \faHandORight \xspace \textbf{Community (E)}
Challenges in this category concern the lack of a community to interact with.
Indeed, many developers wish to have other people they can compare their work with and find support: 12 respondents reported that they suffer from ``\textit{lack of professional connections; lack of peer guidance}''.
These kind of challenges also concern the difficulty of performing code reviews because of missing peers.
In particular, developers pointed out that ``\textit{Code reviews are slow}'', and the issue affects the entire review process.
Furthermore, the additional effort necessary to comprehend the programs does not promote code reviews for developers who want to ``\textit{learn to review the source code to understand how to use some features}''.

\begin{summarybox}[RQ2]
Our results identify challenges related to the comprehension of quantum programs, the hardness of setting up hardware and software infrastructures, the implementation and code quality issues, the difficulty of building a quantum developer's community, and the lack of realism of the current quantum applications.
\end{summarybox}
% !TeX spellcheck = en_US
\section{Discussion and Implications} \label{sec:implications}
Our study's key findings clearly show that quantum programming is still an emerging subject, even though the software engineering approaches available to developers are still limited, and, perhaps more crucially, there are few real-world applications for quantum technology. Both the research questions of the study converge toward a clear implication for the research community: 

\keyfindingsone{There is the need for a joint research effort toward the definition of follow-up empirical studies aiming at delving into the quantum programmers' needs, alongside to the development of tools that these programmers can use to comprehend, set up the required infrastructure, implement, and verify quantum programs. Our study defined a set of research directions to pursue, \revised{represented by the challenges defined in our taxonomy,} along with concrete and specific problems that quantum programmers currently experience and that the research community is called to address.}

While the key take away message is similar to what has been somehow reported in early research on the matter \cite{piattini2021toward,piattini2020talavera,zhao2020quantum,Elaoun2021icsme}, we aimed at elaborating more on this point. 

By nature, survey studies like the one we conducted in this paper might not necessarily provide insights at a granularity level that would allow a proper understanding and analysis of the developer's opinions~\cite{coughlan2009survey}. For instance, some of the open answers might not be clear enough or might even be contrasting. 
For this reason, we decided to complement the insights from the survey with additional, finer-grained observations into the potential adoption of quantum technologies and the role of software engineering for quantum programming. 
To this aim, we created several discussion groups on developer's forums. 
Being these discussions defined \emph{after} the analysis of the survey responses, our goal was to clarify or explore more closely some of the aspects mentioned in the survey rather than creating generic discussions about quantum software engineering. 

We opted for \textsc{Reddit},\footnote{\textsc{Reddit} website: \url{https://www.reddit.com}.} a popular social news aggregation, web content rating, and discussion website that developers often use to discuss open issues and challenges on various computer science-related subjects~\cite{medvedev2017anatomy}. 
In particular, \textsc{Reddit} allows the creation of the so-called \emph{sub-reddits}, \ie discussion channels dedicated to specific matters.
In our case, there exist three sub-reddits dedicated to quantum computing and programming, \eg \texttt{`r/QuantumComputing'}, \texttt{`r/Qiskit'}, and \texttt{`r/cirq'}.
\footnote{At the submission date, there was no sub-reddits specific for \textsc{Q\#}.} 
Hence, also in this case we could target a population of developers that are currently working on quantum programming. 

As already mentioned, the goal of such an additional analysis was to let developers discuss specific aspects that were found to be particularly interesting, contrasting, or unclear from the survey responses.
These pieces of information were extracted from the first author of the paper after the survey analysis.
For each of the identified discussion points, we created a post where we presented ourselves, the goals of the study, and the topic we were interested in.
We asked developers to comment and provide their take on it.
As an example, let consider the case of \revised{theoretical ground}---this is one of the key challenges discovered and detailed in \Cref{sec:taxonomy}.
We created the following post (the introductory part is omitted for the sake of brevity):  

\examplebox{It seems that this documentation, although simplifying most of the concepts related to quantum mechanics, is still really hard to understand, and most of the parts assume a deep knowledge of linear algebra. Is there a way for a beginner who has mainly a computer science background to start programming with quantum technologies?}

As shown, the post aimed to engage developers in a discussion on how to use the documentation of quantum frameworks to start programming with quantum technologies. Similar posts were created for the other discussion points: the complete list of posts is available in our online appendix~\cite{appendix}. Overall, we obtained around 30 answers. 

Among these discussions, we found of particular interest the ones about the real applicability of quantum programming and the difficulty for a practitioner with only a computer science background to cope with these technologies, along with other questions regarding code quality issues. An interesting point of discussion was concerned the lack of abstraction which affect the quantum technologies taken into consideration.
We found out that the idea of programming is completely different, which \emph{``is more like designing circuits rather than programming the high level stuff. Much like probably 60 to 70 years ago when the concept of computers wasn't well defined and those which existed back then were specific purpose computers and making them would require designing circuits (similar to what we are doing in Quantum programming).''}
This comment suggests that the software engineering researchers might take advantage of the methodologies defined in the past, which resulted in the software and abstraction layers that we currently have, to apply them to quantum computers.

As for the background challenge, we found something interesting. While it is generally recognized that to be able to program a quantum computer a knowledge of quantum physics and advanced linear algebra is required, the discussions we had let emerge it as a \emph{misconception} or \emph{false myth} of quantum programming.
This was made clear by one of the developers who reported that: \emph{``Much like how a deep understanding of the band structure of the Silicon in a CPU is not required to do computer science, the end goal of quantum programming is to be able to program on a quantum computer without needing knowledge of what is physically happening to the qubits during the computation''}.
Of course, being familiar with some notion of math can help in having a better understanding of the underlying technology, but is not mandatory to be a quantum physics expert---indeed, many books have been published which propose an ``hands on'' approach, such as the one written by Johnston \etal \cite{johnston2019programming}.

Speaking about \emph{misconceptions}, what emerges from the discussions about the use of quantum programming in real world applications is that many developers struggle to find an actual application of these algorithms.
These discussions lead to the idea that quantum computer might sometime replace digital computers, and developers are adapting their programming abilities to achieve that.
However, what we know now is that a complete replacement of digital computer is not a matter of near future, given several technical limitations.
One of the developers that discussed with us about this topic (using Grover's Algorithm as an example), said that \emph{``real world use cases of Grover's algorithm would definitely take at least thousands of fault tolerant qubits, which at the moment would require millions of noisy qubis'',} which the current state of the art technology cannot give.
What we are really able to do with quantum computer is solving optimization problems exploiting variational quantum algorithms on near terms devices (\ie the quantum computers available so far) \cite{wecker2015progress, moll2018quantum, cerezo2021variational}.
In this scenario, a quantum computer should be seen as an external computational unit (Quantum Computational Unit, QPU), and programmed with that idea in mind, just as Graphical Processing Units (GPUs) are programmed \cite{johnston2019programming}.
Thus, we can claim that all the challenges which we have shown should be solved by software engineering research with this idea in mind, trying to apply the acquired knowledge with GPU programming and all its issues and challenges to this new emerging field.
Remembering that as Booch remarks \cite{booch2018history}: ``No matter the medium or the technology or the domain, the fundamentals of sound software engineering will always apply: craft sound abstractions.''
Without this in mind, \emph{misconceptions} are behind the corner, making us far from software engineering for quantum programming.
\revised{The message for practitioners aiming at joining the QSE challenge derives from this: quantum programming should not be considered the panacea of all evils; its current applications are still minimal, expecially at the industry level.
Practitioners should focus on leveraging the benefits of quantum programming in particular situations.
In doing so, they could develop new software engineering methods and tools to boost the research in the field.}
% !TeX spellcheck = en_US
\section{Threats to Validity} \label{sec:ttv}
A number of threats might have possibly influenced the results of our study. In the following, we report and discuss how we mitigated them.

\subsection{Threats to Construct Validity}
Threats to construct validity concern the relationship between hypotheses and observations. When studying the current adoption of quantum programming (\textbf{RQ$_1$}), we mined repositories using search strategies aiming at identifying all the projects using the most widely used quantum frameworks available to date. These strategies pertain to source code patterns that developers must necessarily use to include the frameworks in their code or, in the case of \qsharp, through the features provided by \textsc{GitHub}. In any case, our mining strategies represent the only available, hence ensuring the maximum coverage possible. In this respect, we can claim that the analysis done is extensive and not threatened by false positive or false negative projects. 

As for \textbf{RQ$_2$}, whenever needed we defined free text answers to let practitioners express their opinions freely, without any restriction. As for the participants involved, we invited developers who contributed to the \textsc{GitHub} projects relying on quantum frameworks. Based on the considerations done for the projects selection, we can ensure that the developers involved in the survey are the ones that are actually working on open-source quantum projects. In addition, we excluded the developers who contributed to those projects with less than ten commits. This was done to make sure to rely on the opinions of developers who had done significant contributions in terms of quantum programming. Nonetheless, replications of the survey study would be beneficial to discover additional points of view and perspectives on the state of quantum programming. 

\subsection{Threats to Conclusion Validity}
A threat of this kind which might affect the validity of our study is related with the subjectivity of the constructed taxonomies, either the one regarding the current adoption of quantum technologies or the one related to the quantum programming challenges, which might result in a biased classification.
In both cases we iteratively built them by splitting and aggregating categories, following a rigorous schemed procedure.
Moreover, different authors have taken part of the taxonomy building phase, which provides more confidence of the result achieved. Anyway, also in this case replications would be desirable. To ease the work of other researchers, we released all the material produced in the context of this study in our online appendix \cite{appendix}. 

\subsection{Threats to External Validity}
These threats concern with the generalization of our results.
\revised{First, the scope of this study is limited to the quantum logic gate model of quantum computing.
On the one hand, we did focus on a single model given the exploratory nature of the study.
On the other hand, follow-up studies on other quantum models, such as the very promising quantum annealing one, are already part of our research agenda.}
The study considers the three state-of-the-practice and most mature quantum technologies available so far~\cite{quantum1,quantum2}, namely \qiskit~\cite{aleksandrowicz2019qiskit}, \cirq~\cite{cirq_developers_2021_4750446}, and \qsharp~\cite{microsoft_quantum}. While other quantum frameworks are emerging, we leave to further research the willingness to compare the results achieved in our empirical study. 
\revised{It must also be considered that our study focuses solely on open-source quantum projects, which can be freely mined and analyzed. Thus the selected repositories only come from \textsc{GitHub} which offers a plethora of mining tools.
We are aware that other hosting platforms are available (\eg Quantum Programming Studio), although not automatically minable; therefore, we acknowledge a threat to validity to the generalizability of our results.
It is worth mentioning that the latter platform is technology-agnostic: its inclusion in our work would have been out of scope since the circuits were not designed for the languages we took into consideration but rather translated. They can only host quantum circuits (\ie algorithms) and not entire quantum applications (which require a classical part).
Finally, our results observations might not be valid in industrial contexts.
We acknowledge a limitation of our study, limited by the proprietary and closed-source nature of industrial code.
}
\revised{The majority of the mined repositories are developed for didactic purposes, which might represent a potential bias to the final results.
This characteristic might have biased the results, especially in the case of the faced challenges, which might pertain mainly to developers who are learning the quantum technologies under consideration.
Therefore, we plan to replicate this study once quantum programming is more mature and pervasive to improve the generalizability of the already achieved results, targeting a different population of developers.
Another potential threat might be represented by the fact that we selected only developers coming from the contributors of the mined \textsc{GitHub} repositories.
On the one hand, this might introduce a bias in the generalizability of the study.
On the other hand, we had to select a very niche set of developers, so we chose stricter selection criteria.}
Another potential threat might be represented by the response rate achieved when involving developers in the survey study. In particular, we got an answer from 56 out of 905 contacted developers, which corresponds to a response rate of 6,1\%. 
Such a response rate is in line with respect to other papers that conducted survey studies in software engineering \cite{blackburn1996improving}: as such, we deemed the response rate to be good enough. Of course, further experimentation might reveal additional insights that were not discussed by the participants of our study. 
\revised{Finally, the selection criteria we adopted (as recommended by Sugar \etal \cite{sugar2014studies}) might have excluded a more diverse and potentially more experienced population of developers.
We are aware of this risk, and we will consider different selection criteria in the follow-up studies.}

% !TeX spellcheck = en_US
\section{Conclusions} \label{sec:conclusion}
We have conducted an exploratory study into the current state of quantum programming. First, we applied a manual coding exercise to understand \revised{for which purposes the currently available software repositories that use quantum technologies are created}. Secondly, we surveyed 56 quantum programmers, inquiring them on the current usage and challenges they face when interacting with the state of the practice quantum frameworks, \ie \qiskit~\cite{aleksandrowicz2019qiskit}, \cirq~\cite{cirq_developers_2021_4750446}, and \qsharp~\cite{microsoft_quantum}. The results of our empirical study revealed that quantum programming is mainly used for didactic purposes or for curiosity to experiment with quantum technologies. In addition, the survey study shed light on several challenges which are not only related to technical aspects of quantum development but also socio-technical matters, \revised{such as the comprehension challenges, the degree of realism, and the community ones}.

The output of the study represents the input of our future research agenda. We indeed aim at addressing the challenges identified, by providing (semi-)automated support for developers in terms of quantum program comprehension, analysis, manipulation, and testing. 

\section*{Acknowledgement}
Fabio gratefully acknowledges the support of the Swiss National Science Foundation through the SNF Projects No. PZ00P2\_186090.

\balance
\bibliographystyle{model1-num-names}
\bibliography{biblio}

\end{document}